\documentclass[nofootinbib,superscriptaddress,prd,twocolumn]{revtex4-2}
\usepackage{amsmath,graphicx,hyperref}
\allowdisplaybreaks

\newcommand{\mc}{\mathcal}
\newcommand{\mr}{\mathrm}

\newcommand{\mO}{\mathcal{O}}
\newcommand{\mQ}{\mathcal{Q}}
\newcommand{\mJ}{\mathcal{J}}

\newcommand{\be}{\begin{equation}} 
\newcommand{\ee}{\end{equation}} 
\newcommand{\bea}{\begin{eqnarray}} 
\newcommand{\eea}{\end{eqnarray}}

\newcommand{\n}{\overline{n}}

\newcommand{\nnb}{\nonumber} 

\newcommand{\as}{\alpha_s}

\DeclareMathOperator{\arccot}{arccot}

\begin{document}

%\vspace*{18pt}

%%%%%%%%%%%%%%%%%%%%%%%%%%%%%%%%%%%%%%%%%%%%%%%%%%%%%%%%%%%%%%%%%%%%%%
%%%%%%%%%%%%%%%%%%%%%%%%%%%%% Title %%%%%%%%%%%%%%%%%%%%%%%%%%%%%%%%%%
%%%%%%%%%%%%%%%%%%%%%%%%%%%%%%%%%%%%%%%%%%%%%%%%%%%%%%%%%%%%%%%%%%%%%%

\title{Large $N$-point energy correlator in the collinear limit}

\def\Seoultech{Institute of Convergence Fundamental Studies and School of Natural Sciences, Seoul National University of Science and Technology, Seoul 01811, Korea}
\def\Pitt{Pittsburgh Particle Physics Astrophysics and Cosmology Center (PITT PACC) \\ Department of Physics and Astronomy, University of Pittsburgh, Pittsburgh, Pennsylvania 15260, USA}
\def\GuangxiU{School of Physical Science and Technology, Guangxi University, Nanning 530004, P. R. China}
\def\GuangxiUKeyLab{Guangxi Key Laboratory for Relativistic Astrophysics, School of Physical Science and Technology, Guangxi University, Nanning 530004, P. R. China}

\author{Lin Dai}
\email[E-mail:]{dailin@gxu.edu.cn}
%\affiliation{\GuangxiU}
\affiliation{\GuangxiUKeyLab}
\author{Chul Kim}
\email[E-mail:]{chul@seoultech.ac.kr}
\affiliation{\Seoultech} 
\author{Adam K. Leibovich}
\email[E-mail:]{akl2@pitt.edu}
\affiliation{\Pitt}

\begin{abstract} 
For the $N$-point energy correlator in the collinear limit, the  largest projected angle $R$ in the large $N$ limit can  be viewed as the radius of the jet that encompasses all the collinear core particles, while contributions from soft gluons to the radius are suppressed. We relate the $N$-point energy correlator in the large $N$ limit to the moments of the fragmentation functions to a jet, and, using previous work, we compare the difference between jets initiated by heavy quarks and light quarks. This comparison demonstrates the deadcone effect.
\end{abstract}

\maketitle 

%%%%%%%%%%%%%%%%%%%%%%%%%%%%%%%%%%%%%%%%%%%%%%%%%%%%%%%%%%%%%%%%%%%%%%

\section{Introduction} 

The jet, a collimated spray of energetic particles, is a characteristic phenomenon in high energy collisions. A fundamental question about the jet is the formation mechanism, which can be gleaned by studying jet substructure.
Recently, interest in the energy-energy correlator~(EEC)~\cite{Basham:1977iq,Basham:1978zq} has been revived, focusing on the jet substructure~\cite{Dixon:2019uzg}. Interestingly, the generalization of the EEC to $N$-points introduces a successful framework to analyze jet substructure in an intuitive and systematic way~\cite{Chen:2020vvp,Komiske:2022enw,Lee:2022uwt,Craft:2022kdo}.

\begin{figure}[h]
\begin{center}
\includegraphics[height=6cm]{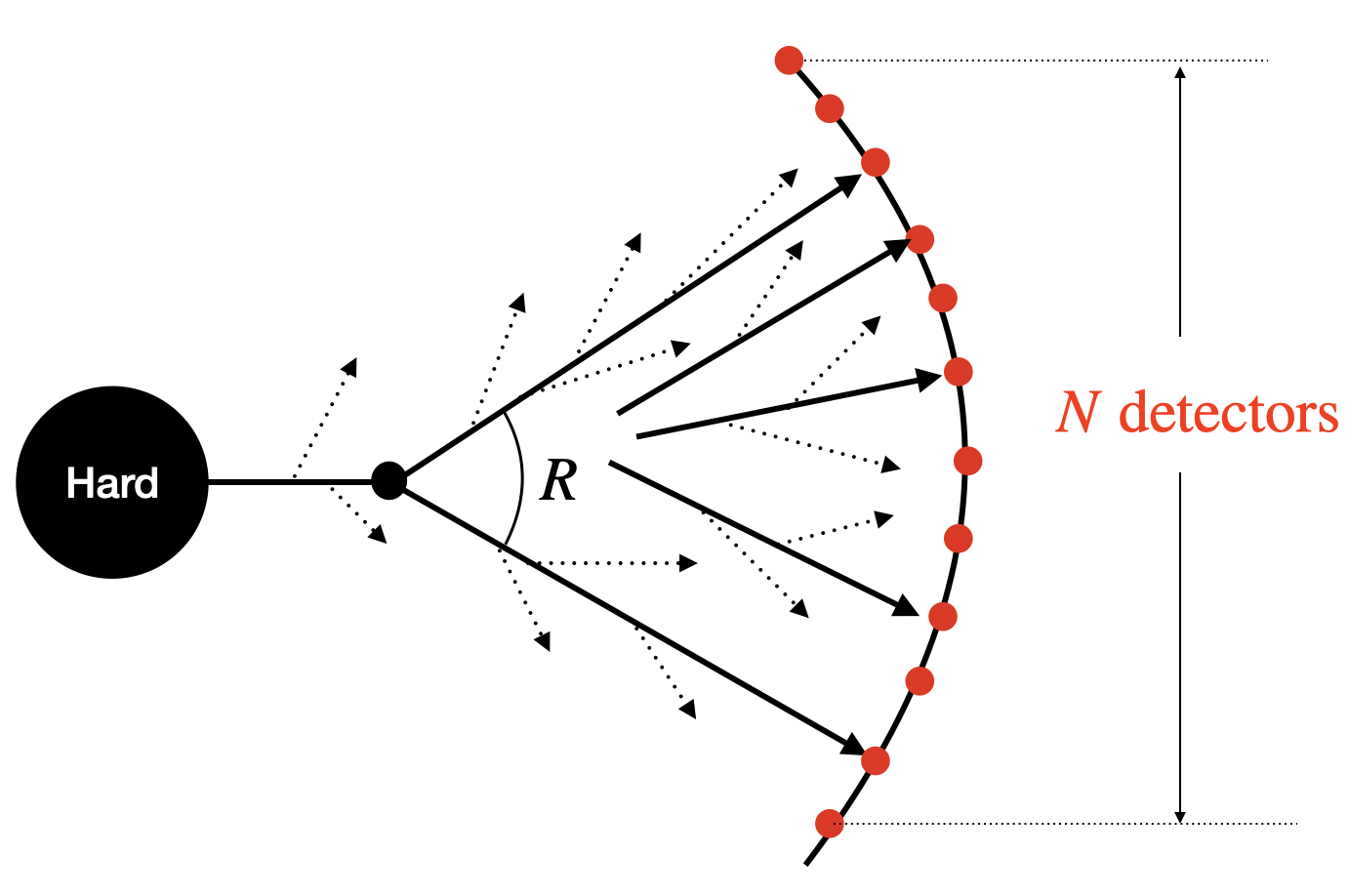}
\end{center}
\vspace{-0.8cm}
\caption{\label{Ndetectors}
Illustration of the (projected) $N$-point energy correlator in the large $N$ limit.
Here thick straight lines denote collinear particles and thin dotted lines are soft or collinear-soft particles. $R$ represents the largest angle between collinear particles. 
}
\end{figure}

When using the $N$-point energy correlator~(ENC) to examine the jet substructure, an interesting question arises: what aspect of the jet can we uncover as $N$ approaches infinity? Essentially, in the limit $N \to \infty$, a large number of detectors would capture a whole bunch of energetic particles, as depicted in Fig.~\ref{Ndetectors}. 
Remarkably, since the ENC is weighted by the energies of the particles, the contributions from soft or collinear-soft~(csoft) gluons\footnote{ 
We shall consider the limit where the projected angle $R$ is much smaller than one. In this case, the nonvanishing soft gluons that are able to resolve the jet boundary at $R$ have the scaling shown in Eq.~\eqref{csofts}. We refer to a degree of freedom with this scaling as a ``collinear-soft''~(csoft) mode. 
}
in the measurement by the $N$ detectors can be safely ignored. (As we will see, the (c)soft contribution in the measurement is suppressed by at least  $1/N$.) Thus, if we project the ENC in the large $N$ limit to the largest angle $R$ between the (detected) collinear particles, we can reconstruct a jet with a minimum range that includes the entire set of collinear core particles. In other words, outside the jet with the angular size $R$, there exist only (c)soft radiations. 

Therefore, the projected angle $R$ for the correlator reveals an important characteristic of a jet.
% \footnote{ Since the soft radiations are suppressed, the ENC with the large $N$ is similar to the groomed jet radius~\cite{Larkoski:2014wba,Kang:2019prh,Caucal:2021cfb,ALICE:2022hyz}, while the ENC could be more advantageous than the groomed results as advocated in Ref.~\cite{Chen:2020vvp}. } 
%Since the soft radiations are suppressed, the ENC with the large $N$ is similar to the groomed jet radius~\cite{Larkoski:2014wba,Kang:2019prh,Caucal:2021cfb,ALICE:2022hyz} in the sense that the effects of the soft gluon radiations are effectively removed. 
%However, as advocated in Ref.~\cite{Chen:2020vvp,Lee:2022uwt}, the ENC could be more advantageous. For example, the ENC can be systematically considered with the operator production expansion (OPE) of the energy flow operators being applied, and it has no complexity or nonperturbative effects due to clustering multiparticles following a certain recombinational jet algorithms. 
%Therefore, the projected angle $R$ provides a characteristic measure of the jet. 
Because soft radiation is suppressed, the ENC in the large-$N$ limit plays a role similar to the groomed jet radius~\cite{Larkoski:2014wba,Kang:2019prh,Caucal:2021cfb,ALICE:2022hyz}, in the sense that the impact of soft gluon radiation is effectively removed. 
At the same time, as emphasized in Refs.~\cite{Chen:2020vvp,Lee:2022uwt}, the ENC offers additional advantages. For example, it can be systematically analyzed using the operator product expansion (OPE) of energy flow operators, and it avoids the algorithmic complications and nonperturbative effects associated with clustering multiple particles through recombination jet algorithms.

In this paper, we analyze the projected ENC in the large $N$ limit within the factorization framework using soft-collinear effective theory (SCET)~\cite{Bauer:2000ew,Bauer:2000yr,Bauer:2001yt,Bauer:2002nz}.
We find that the key components in establishing the factorization theorem are the fragmentation functions to a jet (FFJs)~\cite{Kang:2016mcy,Dai:2016hzf}, and that the ENC can be expressed in terms of the moments of the FFJs. Additionally, leveraging the established results of the FFJs, we will demonstrate a distinctive feature between light and heavy quark jets, namely, the deadcone effect~\cite{Dokshitzer:1991fc,Dokshitzer:1991fd}. 

The paper is organized as follows. In Sec.~\ref{relation}, we show the relationship between the projected ENC and the jet cross section. In Sec.~\ref{largenrefac}, we examine the results in the large $N$ limit, where we are able to resum the large logs of small $R$ and large $N$ to NNL order. Using previous results, we compare the distribution for light quarks and heavy quarks, demonstrating the deadcone effect. In Sec.~\ref{eeENC}, we apply the previous results for the projected ENC for $e^+e^-$ annihilation and compare our results to Pythia. We conclude in Sec.~\ref{conc}.

\section{Relation between the projected ENC and the jet cross section} \label{relation}
To begin with, let us consider the $N$-point energy correlator, 
\begin{align}
   &\frac{d\Sigma}{dz_1\cdots dz_{M_N}} = \sum_{a_1\cdots a_N} \int d\sigma \frac{E_{a_1}\cdots E_{a_N}}{Q^N} \\ 
   &~~~~\times \delta\bigl(z_1-\sin^2\frac{\theta_{12}}{2}\bigr)\cdots 
   \delta\bigl(z_{M_N}-\sin^2\frac{\theta_{N(N-1)}}{2}\bigr), \nnb
\end{align} 
where, in case of $e^+e^-$-annihilation, $Q$ is the center of mass energy,  $M_N = N(N-1)/2$, and $z_k$ run over angles between $N$ detectors. 
Then, the projected correlator to the largest angle~\cite{Chen:2020vvp} is given by 
\be 
\frac{d\Sigma}{dz_L} = \int dz_1\cdots dz_{M_N} \delta\bigl(z_L-z_{\rm max}\bigr)\frac{d\Sigma}{dz_1\cdots dz_{M_N}}\ ,   
\ee
and the cumulant to the angle $R$ is 
\be
\label{defcum}
\Sigma_N (R) = \int^{\sin^2(R/2)}_0 dz_L \frac{d\Sigma}{dz_L}\ .  
\ee

In the collinear limit with $R \ll 1$, this cumulant describes the production of a jet with radius $R$. 
Here, the jet represents a collection of particles enclosed within the radius $R$. 
Thus, this is not a usual jet defined by a recombination algorithm such as C/A~\cite{Dokshitzer:1997in}. 
However, this newly defined jet for the projected ENC (hereafter referred to as the ENC jet) is also infrared (IR) safe, 
and collinear factorization into fragmentation functions~\cite{Collins:1981uw} remains applicable. 
Therefore, the inclusive jet cross section associated with Eq.~(3) can also be expressed in factorized form using the FFJs,\be 
\label{jetxs}
\frac{d\sigma}{dx_J}  = \sigma_0 \sum_{k=q,\bar{q},g} \int^1_{x_J} \frac{dx}{x} H_k \bigl(\frac{x_J}{x},Q,\mu\bigr) D_{J/k} (x,ER,\mu).    
\ee
Here, $\sigma_0$ is the Born cross section, and $x_J = 2E_J/Q$, where $E_J$ is the jet energy. 
The energy of the mother parton $k$ that initiates the jet is given by $E = (x_J/x)\, Q/2$, and 
$x = E_J/E$ denotes the fraction of the jet energy over the mother parton $k$.  
$H_k$ are the hard functions at scale $\sim Q$, and the $D_{J/k}$ are the FFJs (for the ENC jet).

In calculating the ENC FFJs, if two particles are combined into the jet, 
the condition for their angular separation becomes the same as in ${\rm k_T}$-type recombination algorithms,
\be
\theta < R\ . 
\ee
Thus, to the next-to-leading order (NLO) in $\as$, we obtain the same result as the FFJs for the ${\rm k_T}$-type algorithm~\cite{Dai:2016hzf}. 

The FFJs employed in Eq.~\eqref{jetxs} are normalized to satisfy the momentum conservation sum rule,
\be 
\label{msumrule}
\int_0^1 dx\, x\, D_{J/k} (x,ER,\mu) = 1.
\ee
As discussed in Ref.~\cite{Dai:2016hzf}, the argument $ER$, 
which appears in logarithmic terms as $\ln(\mu/ER)$, 
must remain fixed under the integration over $x$ to preserve the momentum sum rule in Eq.~\eqref{msumrule}. Moreover, the factorization structure in Eq.~\eqref{jetxs} 
with $ER$ appearing in the FFJs provide a consistent framework for resumming global logarithms, 
at least up to next-to-leading logarithmic (NLL) accuracy. 
For recent studies of jet cross-section factorization using FFJs at NLL accuracy, we refer to Ref.~\cite{Lee:2024icn}.

In order to obtain the cumulant $\sum_N (R)$ in Eq.~\eqref{defcum} from the factorization of the jet cross section, 
we need to consider the more differential jet cross section that depends on jet substructure variables,  
\begin{align}
   \frac{d\sigma}{dx_J dy_1 \cdots dy_N} 
   &=\sigma_0 \sum_{k} \int^1_{x_J} \frac{dx}{x} H_k \bigl(\frac{x_J}{x},Q,\mu\bigr)   
   \nnb \\
   \label{jetxc1}
   &\times \mc{G}_{J/k} (x,y_1,\cdots, y_N,ER,\mu),  
\end{align}
where $y_i = E_i/E_J$ are the energy fractions of $N$-points over the jet energy. When compared with Eq.~\eqref{jetxs}, we have employed the inclusive jet substructure function $\mc{G}_{J/k}$~\cite{Kang:2016ehg} instead of the FFJs. 
Note, applying the momentum conservation sum rule, we have the relation
\be
\label{gsum}
D_{J/k} (x,ER) =  \int \{ dy_i \}_N \cdot \{y_i\}_N \cdot \mc{G}_{J/k} (x,y_1,\cdots, y_N, ER), 
\ee
where $\{y_i\}_N$ represents $\Pi_{i=1}^N y_i$. 

If we use variables $x_i = 2E_i/Q = y_i x_J$, the differential cross section of Eq.\eqref{jetxc1} is rewritten as $d\sigma/(dx_J\{ dy_i \}_N)= x_J^N \cdot d\sigma/(dx_J\{ dx_i \}_N)$, hence the $N$-point jet cross section is 
\begin{align} 
\frac{d\sigma (R)}{\{dx_i\}_N} &= \int^1_0 \frac{dx_J}{x_J^N} \sum_{k} \int^1_{x_J} \frac{dx}{x} H_k \bigl(\frac{x_J}{x},Q,\mu\bigr)  \nnb \\
\label{npxs}
&~~~\times \mc{G}_{J/k} (x,y_1,\cdots, y_N,ER,\mu). 
\end{align}
%In Refs.~\cite{Dai:2016hzf,Dai:2018ywt}, we  introduced the factorization theorem for the jet cross section with a measured hadron or subjet inside the jet. A similar factorization can be applied for obtaining the factorization theorem of the energy correlator in the collinear limit. If we try to measure energy fractions of $N$ particles (or points) inside a jet of radius $R$, the jet cross section can be written as 
%\begin{align}
%   \frac{d\sigma}{dx_J dy_1 \cdots dy_N} &=  x_J^N\frac{d\sigma}{dx_J dx_1 \cdots dx_N} \nnb \\
%   \label{jetxc1}
%   &=\sum_{\ell} \Bigl(\frac{d\sigma}{dx_J} \Bigr)_{\ell} \cdot \Phi_{\ell} (y_1,\cdots, y_N, R). 
%\end{align}
%$y_i = E_i/E_J$ are the energy fractions over the jet energy, and $x_i = 2E_i/Q$ is given by $x_i = x_J y_i$. $\ell$ denotes the initial parton that starts the formation of the jet, and $\Phi_{\ell}$ is the scale invariant jet substructure function, which satisfies the momentum conservation sum rule 
%\be 
%\label{sumrule}
% \int \{ dy \}_N \cdot \{y\}_N  \cdot \Phi_{\ell} (y_1,\cdots, y_N, R) = 1, 
%\ee
%where $\{y\}_N$ represents $\Pi_{k=1}^N y_k$. 
Therefore, using Eqs.~\eqref{gsum} and \eqref{npxs}, 
%From Eq.~\eqref{jetxc1}, the $N$-point jet cross section is 
%\be 
%\frac{d\sigma (R)}{\{dx\}_N}= \int \frac{dx_J}{x_J^N} \sum_{\ell} \Bigl(\frac{d\sigma}{dx_J} \Bigr)_{\ell} \cdot \Phi_{\ell} (y_1,\cdots, y_N, R).
%\ee
the cumulant in Eq.~\eqref{defcum} can be expressed as 
\begin{align}
    \Sigma_N (R) &= \frac{1}{2^N} \int \{dx_i\}_N \cdot\{x_i\}_N \cdot \frac{d\sigma (R)}{\{dx_i\}_N}  \\
    \label{fact1}
    &= \frac{1}{2^N}\int^1_0 dx_J x_J^N  \sum_{k} \int^1_{x_J} \frac{dx}{x} H_k \bigl(\frac{x_J}{x},Q,\mu\bigr)
     \\
    &\times \int \{dy_i\}_N \cdot\{y_i\}_N \cdot \mc{G}_{J/k} (x,y_1,\cdots, y_N,ER,\mu) \nnb \\
    &= \frac{1}{2^N} \int^1_0 dx_J x_J^N \Bigl(\frac{d\sigma}{dx_J}\Bigr)\ . 
\end{align}    
%Here, in the second line of Eq.~\eqref{fact1}, we applied the momentum sum rule in Eq.~\eqref{sumrule}. Then $\sum_l (d\sigma/dx_J)_{\ell}$ becomes $d\sigma/dx_J$ with the inclusive FFJs, $D_{J/k} = \sum_{\ell} D_{J_{\ell}/k}$, employed. 
Thus, the cumulant ends up as the moments of the inclusive jet cross section.

Using the factorization form of the jet cross section in Eq.~\eqref{jetxs}, the cumulant in factorized form becomes\footnote{
If we consider hadron collision, $H_k$ include the parton distribution functions and $ER$ becomes $p_T R$, where $p_T$ is the large transverse momentum to the beam direction and $R$ is given as $R = \sqrt{\Delta y^2 + \Delta \phi^2}$.
} 
\begin{align}
 \Sigma_N (R) &= \frac{\sigma_0}{2^N} \sum_k \int^1_0 d\zeta \,\zeta^N H_k (\zeta, Q,\mu) \nnb \\
 \label{cumfact}
 &\times \bar{J}_{k,\rm{ENC}} (N,\zeta QR/2,\mu),
\end{align}
where $\zeta = x_J/x= 2E/Q$, and $\bar{J}_{k,\rm{ENC}}$ is the ENC jet function which describes the collinear part of the jet. This result, Eq.~\eqref{cumfact}, from the factorization of the jet cross section, is consistent with the factorization result using ``the $\nu$-correlator" proposed in Ref.~\cite{Chen:2020vvp}. 

The new thing in this paper is that the ENC jet functions $\bar{J}_{k,\rm{ENC}}$ are given by the $N$-th moments of the FFJs,
\begin{align}
\label{ENCjetf}
    \bar{J}_{k,\rm{ENC}} (N,ER,\mu) = \int^1_0 dx\, x^N D_{J/k} (x,ER,\mu). 
\end{align}
We therefore can easily obtain $\bar{J}_{k,\rm{ENC}}$ from the calculation of the FFJs. 
(The leading results in $\as$ is trivially given by $\bar{J}_{k,\rm{ENC}}^{(0)} =1$.)
For example, using the one loop results of the FFJs in Ref.~\cite{Dai:2016hzf} in the massless limit we obtain the results of the EEC jet functions with $N=2$: 
\begin{align}
\label{JqEEC}
    \bar{J}_{q,\rm{EEC}}^{(1)} &= \frac{\as C_F}{4\pi} \Bigl(-3 \ln \frac{\mu^2}{E^2R^2} - \frac{37}{3} \Bigr), \\ 
    \label{JgEEC}
     \bar{J}_{g,\rm{EEC}}^{(1)} &= -\frac{\as}{4\pi} \Bigl(\frac{14C_A+n_f}{5}  \ln\frac{\mu^2}{E^2R^2}+ \frac{898C_A+ 
42 n_f}{75} \Bigr), 
\end{align}
where $n_f$ is a number of quark flavors. These results agree with Ref.~\cite{Dixon:2019uzg}. 

\begin{widetext}
For heavy-quark (HQ) production, using the HQ FFJ from Ref.~\cite{Dai:2018ywt}, we can immediately obtain the HQ ENC jet function. The NLO contribution to the HQ FFJ is given as 
\begin{align}
\label{HQFFJ}
D_{J/\mQ}^{(1)} (x,ER,m,\mu) &= \frac{\as C_F}{2\pi} \Biggl[\frac{1+x^2}{1-x}\Bigl(\ln\frac{\mu^2}{E^2R^2(x^2+b)}-2\ln(1-x) \Bigr)-\frac{2x}{1-x} \frac{b}{x^2+b} -(1-x)\Biggr]_+ \\
&+\frac{\as C_F}{2\pi} \Biggl(\frac{1+(1-x)^2}{x}\Bigl(\ln\frac{\mu^2}{E^2R^2((1-x)^2+b)}-2\ln x \Bigr)-\frac{2(1-x)}{x} \frac{b}{(1-x)^2+b} -x\Biggr)\ , \nnb 
\end{align}
where $b \equiv m^2/(E^2R^2)$, $m$ is the heavy quark mass, and $[\cdots ]_+$ denotes the standard plus distribution.
For $N=2$, we take the second moments of $D_{J/\mQ}^{(1)}$,  obtaining the NLO contribution to the HQ EEC jet function  
\begin{align}
    \bar{J}^{(1)}_{\mQ,\rm{EEC}} (ER,m,\mu) &%= \int^1_0 dx x^2 D^{(1)}_{J/\mQ}(x,ER,m,\mu) 
    \label{JQEEC}
    = \frac{\as C_F}{4\pi} \Biggl(-3 \ln \frac{\mu^2}{B^2} -\frac{37}{3} +9b + b(2-b) \ln\frac{1+b}{b} -8b^{3/2} \arccot{\sqrt{b}}  
   \Biggr)\ , 
\end{align}
where $B\equiv\sqrt{E^2R^2+m^2}$. If we take the limit $m\to 0~(b\to 0)$, we recover the result of the light-quark case, Eq.~\eqref{JqEEC}. For top-quark production, we can also consider the inverse limit, $m\gg ER$. In this case, Eq.~\eqref{JQEEC} can be expressed as the moments of the standard HQ fragmentation function~\cite{Mele:1990cw} while the $R$ dependence becomes suppressed at higher order in powers of $E^2R^2/m^2$. 
\end{widetext}

\section{Refactorization in the large $N$ limit} \label{largenrefac}

In the large $N$ limit, contributions with large $\zeta$ and $x$ in Eqs.~\eqref{cumfact} and \eqref{ENCjetf} dominate. In this case, the jet takes most of energy from the mother parton $k$~($E_J \sim E$), $x$ approaches 1, and only (c)soft gluons radiate outside of the jet. Additionally, in the region $x\to 1$, the FFJs can be further factorized into the collinear and the csoft pieces~\cite{Dai:2017dpc},
\begin{align}
\label{facFFJs}
    D_{J/k} (x\to 1, ER,\mu) %\approx D_{J_k/k} (x\to 1, E_JR,\mu) \nnb \\
    &= \mc{J}_k(E_J R,\mu) S_{k} (x, E_JR,\mu).
\end{align}
When we write the jet as a sum over contributions from initiating partons $\ell$, $J=\sum_{\ell} J_{\ell}$,
we can ignore the transition to a different flavor, i.e., $k\to J_{\ell\neq k}$ since it is suppressed by $1-x$. $\mc{J}_k$ are the integrated jet functions for collinear particles that radiate within the angle $R$~\cite{Cheung:2009sg,Ellis:2010rwa,Chay:2015ila}, and $S_k$ are the csoft functions responsible for csoft gluon radiations~\cite{Dai:2017dpc}. 
Hence the ENC jet function with large $N$ factorizes,
\be 
\label{JENCfact}
 \bar{J}_{k,\rm{ENC}} (N,E_J R,\mu)=  \mc{J}_k(E_J R,\mu) \bar{S}_k (N, E_J R, \mu),  
\ee
where $\bar{S}_{k}(N)$ are the $N$-th moments of $S_{k}(x)$ in the large $N$ limit.
The scaling behaviors for the collinear ($\mJ_k$) and the csoft ($\bar{S}_k$) momenta are given by 
\begin{align}
    (\n\cdot p_c,p_c^{\perp},n\cdot p_c )  &\sim E_J (1, R,R^2),\\
    \label{csofts} 
    (\n\cdot p_{cs},p_{cs}^{\perp},n\cdot p_{cs} )  &\sim \frac{E_J}{N} (1, R,R^2),
\end{align}
where we have introduced the lightcone vectors satisfying $n^2=\bar{n}^2 = 0$ and $n\cdot \n=2$. Here the jet is described as moving in the $n$-direction. 

Since the energy of the csoft gluon is suppressed by $1/N$, the measurement of the csoft gluon radiations (by $N$ detectors) is power-counted by at most $\mO(1/N)$. So, at leading order in the large $N$ limit, the csoft radiations are present only inclusively and assigned to the csoft function $\bar{S}_k$, while the collinear radiations within the $N$ detectors are described by $\mJ_k$ with the momentum sum rule in Eq.~\eqref{gsum} being applied. 

%Using the heavy quark~(HQ) FFJs~\cite{Dai:2018ywt}, we can immediately obtain the HQ ENC jet function. For $N=2$, the next-to-leading order~(NLO) result of the EEC jet function is given by 
%\begin{widetext}
%\begin{align}
%    \bar{J}_{\mQ,\rm{EEC}} (ER,m,\mu) &= \int^1_0 dx x^2 \Bigl[D_{J_{\mQ}/\mQ}(x,ER,m,\mu) + D_{J_{g}/\mQ} (x,ER,m,\mu) \Bigr] \nnb \\
%    \label{JQEEC}
%    &= 1+ \frac{\as C_F}{4\pi} \Bigl[-3 \ln \frac{\mu^2}{B^2} -\frac{37}{3} +9b + b(2-b) \ln\frac{1+b}{b} -8b^{3/2} \arccot{\sqrt{b}}  
%   \Bigr]\ , 
%\end{align}
%\end{widetext}
%where $B\equiv\sqrt{E^2R^2+m^2}$ and $b\equiv m^2/E^2R^2$. If we take the limit $m\to 0~(b\to 0)$, we recover the result of the light quark case in Eq.~\eqref{JqEEC}. We can also consider the inverse limit, $m\gg ER$. In this case, Eq.~\eqref{JQEEC} can be expressed as the moments of the standard HQ fragmentation function~\cite{Mele:1990cw} while the $R$-dependence becomes suppressed as the higher order in powers of $E^2R^2/m^2$. 
%\footnote{
%We can also consider the inverse limit, $m\gg E_JR$. In this case, Eq.~\eqref{JQEEC} can be expressed as the moments of the standard heavy quark fragmentation function~\cite{Mele:1990cw} while the $R$-dependence becomes suppressed as the higher order in powers of $E^2R^2/m^2$.   
%}

\begin{widetext}
In the large $N$ limit, the HQ FFJ can be also factorized like Eq.~\eqref{facFFJs}, hence the ENC jet function is given by the multiplication of the integrated HQ jet function and the HQ csoft function in this limit (i.e, $\bar{J}_{\mQ,\rm{ENC}} = \mJ_{\mQ} \bar{S}_{\mQ} (N)$). 
The one loop results are~\cite{Dai:2021mxb} 
    \begin{align}
    \label{HQIJFnlo} 
        \mJ^{(1)}_{\mQ} (E_JR,m,\mu) &= \frac{\as C_F}{4\pi} \Biggl[ \frac{3+b}{1+b} \ln\frac{\mu^2}{B^2} +\ln^2 \frac{\mu^2}{B^2}+ 2f(b)+2g(b) 
        + \frac{4+2\ln(1+b)}{1+b}
-\ln^2 (1+b) -2\mr{Li}_2 (-b) + 4 -\frac{\pi^2}{6} \Biggr]\ ,
         \\
         \label{mSQnlo}
\bar{S}^{(1)}_{\mQ}(N,E_JR, m,\mu) &=\frac{\as C_F}{4\pi}\Bigl[\frac{2b}{1+b} \ln \frac{\mu^2\bar{N}^2}{B^2}-\ln^2\frac{\mu^2\bar{N}^2}{B^2} -\frac{2\ln (1+b)}{1+b}  +\ln^2(1+b)+2\mr{Li}_2 (-b)-\frac{\pi^2}{2}\Bigr]\ , 
   \end{align}
\end{widetext}
where $\bar{N}\equiv N e^{\gamma_E}$. 
$f(b)$ and $g(b)$ are given by the integral forms, 
\begin{align}
\label{fb}
f(b) &= \int^1_0 dz \frac{1+z^2}{1-z} \ln\frac{z^2+b}{1+b}, \\
\label{gb}
g(b) &= \int^1_0 dz \frac{2z}{1-z}\Bigl(\frac{1}{1+b}-\frac{z^2}{z^2+b}\Bigr).  
\end{align}
These functions have the following limits: $f(0) =5/2-2\pi^2/3$ and $f(\infty)=g(\infty)=g(0)=0$. 
When we take the limit as $m\to 0$ in Eqs.~\eqref{HQIJFnlo} and \eqref{mSQnlo},
we once again obtain the massless NLO result. 
%~($\bar{J}_{q,\rm{ENC}} = \mJ_q \cdot \bar{S}_q (N)$) . 

Through the factorized results for the ENC jet functions in Eq.~\eqref{JENCfact}, we are able to  resum both the large logarithms with small $R$ and large $N$ to NLL accuracy.  To resum, we evolve $\mJ_k~(\bar{S}_k)$ from the scale $\mu$ to its characteristic scale $\mu_j~(\mu_s)$, where the scales have been set as $\mu_j = B$ and $\mu_s = B/\bar{N}$ for jets initiated by a heavy quark while $\mu_j = E_J R$ and $\mu_s = E_J R/\bar{N}$ in case of light quark/gluon.
Given that the (hard-)collinear gluons enter into the restricted phase space within $R$, there arise large nonglobal logarithms~(NGLs)~\cite{Dasgupta:2001sh} at next-to-leading logarithmic (NLL) accuracy. 
The impact of NGLs can be nonnegligible in the large $N$ limit.\footnote{
The form of the NGLs is roughly given by $\ln (\mu_j/\mu_{s}) \sim \ln N$. 
For light quark, the resummation of the NGLs with $\ln N$ has been considered in the factorization of the FFJs in the large $z$ limit~\cite{Dai:2017dpc}.}
This issue will be discussed in future works. 
%Due to the hierarchy between the collinear and the csoft interactions, there may be contributions from nonglobal logarithms~\cite{Dasgupta:2001sh} at next-to-leading logarithmic (NLL) accuracy. This issue will be discussed in future works. 

After simultaneous resummation of large logarithms of small $R$ and large $N$, taking the derivative with respect to $R$ of the ENC jet function, we obtain the resummed contribution to the projected ENC. 
At NLL accuracy, it can be expressed as 
\begin{widetext}\begin{align}
\label{dJdR}
 \frac{d \bar{J}_{k,\rm{ENC}}(\mu)}{dR} =& e^{\mc{M}_N(\mu,\mu_j,\mu_s)}\Bigl[\frac{dM_N(\mu_j,\mu_s)}{dR} \bar{J}^{\rm{NL}}_{k,\rm{ENC}}(\mu_j,\mu_s) +\frac{d\bar{J}^{\rm{NL}}_{k,\rm{ENC}}(\mu_j,\mu_s)}{dR}  \Bigr], 
\end{align}
where $M_N$ is the exponentiation factor for the resummation of the large logarithms and $\bar{J}^{\rm{NL}}_{k,\rm{ENC}}$ are the nonlogarithmic terms in the ENC jet functions at the fixed order $\as$. 

When a heavy quark initiates the jet, the analytical results of $M_N$ and $dM_N/dR$ are given by   
    \begin{align}
    \label{MNQ1} 
       M_N^{\mQ} (\mu,\mu_j,\mu_s) &= -2 S_{\Gamma} (\mu_j,\mu_s)-\ln\frac{\mu_j^2}{B^2}\cdot a_{\Gamma}(\mu_j,\mu_s) -\ln\bar{N}^2\cdot a_{\Gamma}(\mu,\mu_s) -\frac{C_F}{\beta_0} \Bigl(\frac{3+b}{1+b}\ln\frac{\as(\mu)}{\as(\mu_j)}+\frac{2b}{1+b}\ln\frac{\as(\mu)}{\as(\mu_s)}\Bigr),
         \\
         \label{MNQ2} 
\frac{dM_N^{\mQ}(\mu_j,\mu_s)}{dR} &=\frac{2E_J^2R}{B^2}\Biggl[
a_{\Gamma} (B,B/\bar{N}) + \frac{C_F}{\beta_0} \frac{2b}{1+b}\ln\frac{\as(B)}{\as(B/\bar{N})} 
-C_F\Bigl(\frac{\as(B)}{4\pi} \frac{3+b}{1+b} +\frac{\as(B/\bar{N})}{4\pi} \frac{2b}{1+b} \Bigr)
\Biggr]\ . 
   \end{align}
\end{widetext}
In obtaining Eq.~\eqref{MNQ1}, we first set $\mu_j = B$ and $\mu_s = B/\bar{N}$, then take the derivative. 
In Eqs.~\eqref{MNQ1} and \eqref{MNQ2}, $S_{\Gamma}$ and $a_{\Gamma}$, the evolution factors for the cusp anomalous dimensions $\Gamma_C$~\cite{Korchemsky:1987wg}, are given by
\begin{align}
    S_{\Gamma}(\mu_1,\mu_2) &= \int^{\mu_1}_{\mu_2} \frac{d\mu}{\mu} \Gamma_C(\as) \ln\frac{\mu}{\mu_1}\ ,  \\
    a_{\Gamma}(\mu_1,\mu_2) &= \int^{\mu_1}_{\mu_2} \frac{d\mu}{\mu} \Gamma_C(\as)\ .  
\end{align}

\begin{figure}[h]
\begin{center}
\includegraphics[height=5.5cm]{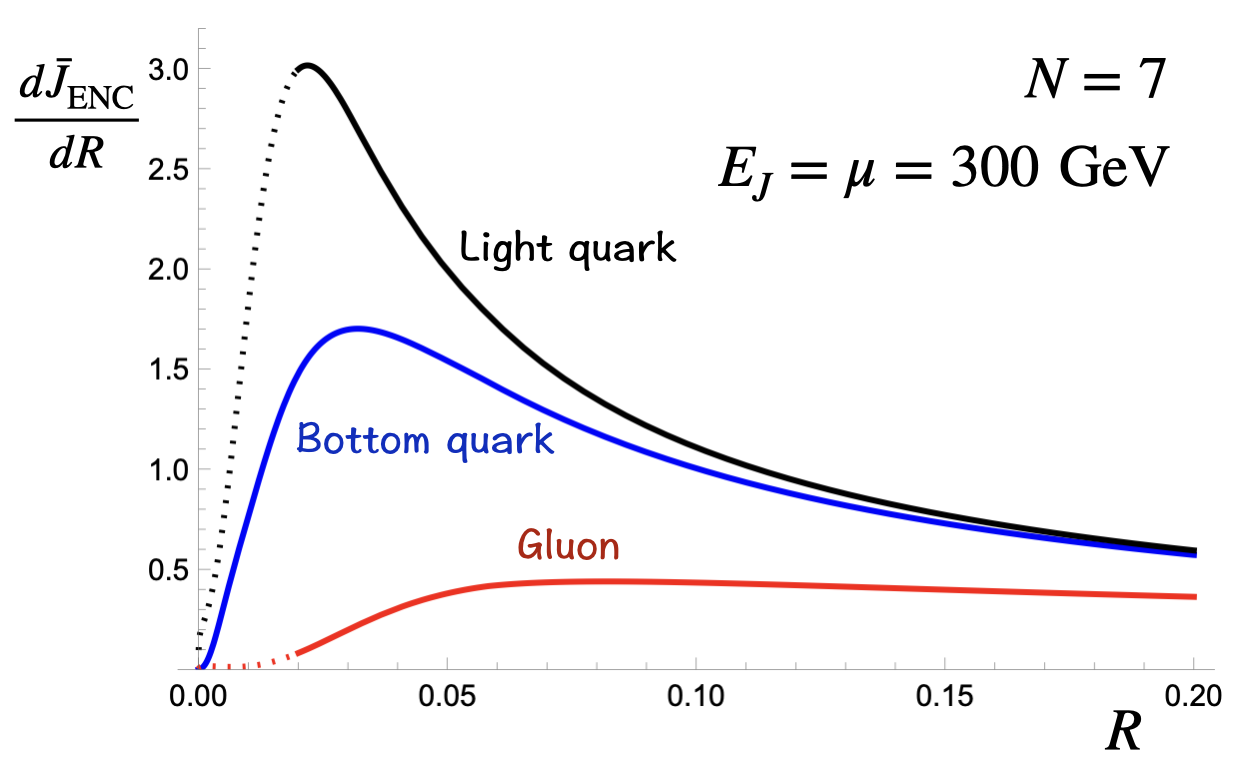}
\end{center}
\vspace{-0.6cm}
\caption{\label{JENC}
Resummed results for the derivatives of the ENC jet functions to NLL accuracy. Here the dotted line for the light quark/gluon represents the nonperturbative domain, while the entire region for the bottom quark distribution can be described perturbatively.  
}
\label{fig:JENC}
\end{figure}

In Fig.~\ref{JENC}, we illustrate the differential distributions for the ENC jet functions with $N=7$. Compared to the light quark distribution, the bottom quark distribution is significantly suppressed in the small $R$ region due to the deadcone effect. 
Near $R=0$, our description of the light quark/gluon distribution 
using a scale profile\footnote{
In order to avoid the Landau pole in case of the light quark/gluon, we have modified the scales $\mu_j$ and $\mu_s$ below some small point $R_0$ in the following way
\begin{align}
    \mu_j(R<R_0) =  \frac{E_J^2}{4\mu_0} R^2 + \mu_0,~~\mu_s(R<R_0) =\frac{\mu_j(R<R_0)}{\bar{N}}. 
\end{align} 
Here we choose $R_0 = 2\mu_0/E_J$, and, in Fig.~\ref{JENC} with $E_J = 300~\rm{GeV}$, the freezing scale $\mu_0$ has been chosen to be $4.5~\rm{GeV}$~(8~GeV) for a light quark~(gluon). 
}
is not exact due to nonperturbative effects. For a precise prediction in this region, comparison with the experimental data is required, whereas the distribution for the bottom quark can be described perturbatively over essentially the entire region due to the heavy quark mass.

\section{The projected ENC in $e^+e^-$-annihilation}\label{eeENC}

Using the resummed results of the ENC jet functions, we next consider the ENC for $e^+e^-$-annihilation in the large $N$ limit. In this case the hard function $H_k$ in Eq.~\eqref{cumfact} can be additionally factorized into the virtual hard function and the inclusive jet function in the  direction opposite to our measured ENC jet. As a result, applying Eq.~\eqref{JENCfact} to Eq.~\eqref{cumfact}, we obtain the factorization theorem of the cumulant for the ENC 
\begin{align}
    \bar{\Sigma} (R) \equiv \frac{2^N}{\sigma_0} \Sigma_N (R) &= \sum_{k=q,\mQ} H_k^v (Q,\mu) \mJ_k (QR/2,\mu)  \nnb \\
    \label{factlc}
    &\hspace{-0.5cm}\times \bar{J}^{\n}_{k}(Q,N,\mu) \bar{S}_k (QR/2,N,\mu),  
\end{align}
where $E_J$ has been replaced with $Q/2$, and we ignore the contribution from jets initiated by gluons since they are suppressed by $1/N$ and $\as$. One loop results for the virtual hard function and the inclusive jet function in the opposite direction are given by 
\begin{align}
  H_{q,\mQ}^{v(1)} (Q,\mu) &= \frac{\as C_F}{2\pi} \Bigl(-3\ln\frac{\mu^2}{Q^2} - \ln^2\frac{\mu^2}{Q^2} -8+\frac{7\pi^2}{6} \Bigr), \\
  \bar{J}_{q,\mQ}^{\n (1)} (Q,N,\mu) &= \frac{\as C_F}{2\pi} \Bigl(\frac{3}{2} \ln\frac{\mu^2 \bar{N}}{Q^2} +\ln^2\frac{\mu^2 \bar{N}}{Q^2} +\frac{7}{2} - \frac{\pi^2}{3}\Bigr). 
\end{align}
Here, for both the heavy ($k=\mQ$) and light quark ($k=q$), we can use the same result, as long as $Q$ and $Q/\bar{N}^{1/2}$ are both taken to be much larger than the heavy quark mass. 

\begin{figure}[h]
\begin{center}
\includegraphics[width=8.5cm]{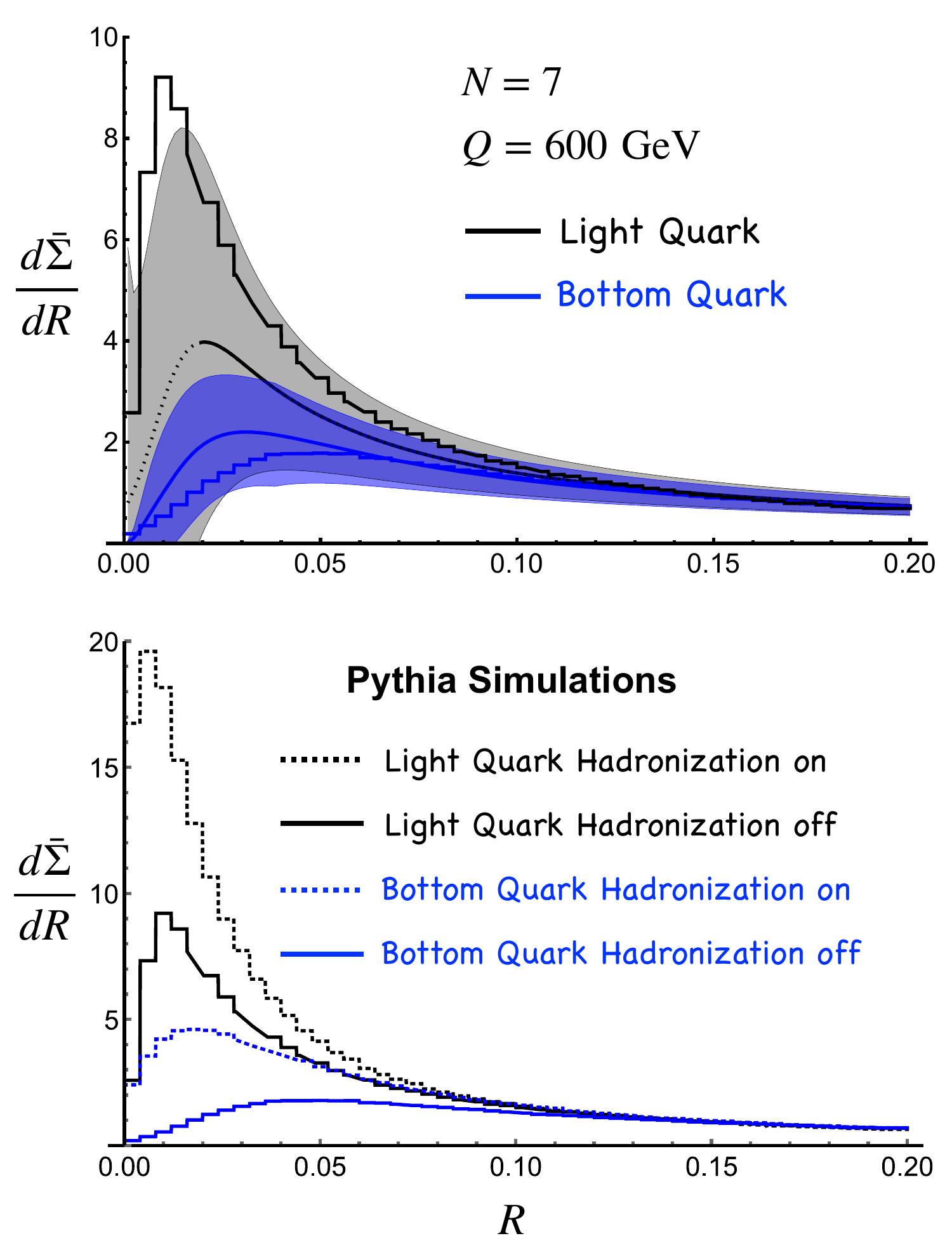}
\end{center}
\vspace{-0.6cm}
\caption{\label{LCENC}
The bottom and the light quark ENCs in $e^+e^-$-annihilation. In the upper panel, the blue and black solid smooth curves (associated with the blue and grey bands, respectively) are from analytic computations at NLL accuracy, and the step curves are from Pythia simulations. The lower panel shows the comparison between Pythia simulations with hadronization turned on and off. 
}
\label{fig:eplus-eminus}
\end{figure}

In Fig.~\ref{LCENC}, we show the $N=7$ ENCs for the bottom quark (the blue solid curve and blue band) and a light quark (the black solid and dotted curves, and the corresponding gray band) in $e^+e^-$ annihilation at NLL accuracy, with collision energy $Q=600 ~\rm GeV$.
Here the bands arise from varying the characteristic scales $\mu_i$ of the factorized functions in Eq.~\eqref{factlc} between $\mu_i/2$ and $2\mu_i$. For the light quark, the errors become large in the  nonperturbative region (the dotted line) from these scale variations.\footnote{
For the recent study of the nonperturbative effect on the light quark ENC, we refer to Ref.~\cite{Lee:2024esz,Chen:2024nyc}.
}  
%as it is too large and covers most of the bottom quark band. 
%Also, since the jet scale is equal to $\sqrt{E_J^2R^2+m^2}$, the light quark jet scale becomes non-perturbative as $R \to 0$. 
This is in drastic contrast with the case of the bottom quark jet, for which the jet scale $\sqrt{E_J^2R^2+m^2}$ is perturbative even when $R$ approaches $0$, and so we are able to have a better perturbative estimate for the heavy quark.  We also have found that the resummed results of the bottom quark ENCs are stable for a wide variation of $N$ (e.g., $N \in [5,10]$).\footnote{
Note, however, that the csoft scale becomes nonperturbative for large $N$ near $R=0$ since the csoft scale,  $B/\bar{N} \approx  m/\bar{N}$, approaches zero.
In Figs.~\ref{fig:JENC} and \ref{fig:eplus-eminus}, 
we have chosen $N$ to be $7$, which corresponds to $m/\bar{N} \sim 0.4$ GeV, and all relevant scales (hard, jet, and csoft) are still above $\Lambda_{\rm QCD}$.
} In Fig.~\ref{LCENC}, we also show the comparison between our analytic calculations and Monte Carlo simulations using Pythia. The step curves in the upper panel shows Pythia simulations with hadronization turned off, which are comparable to our analytic results. The lower panel shows the comparison between Pythia simulations with hadronization turned on (solid) and off (dotted). Whether hadronization is turned on or off, high contrast can be seen between the ENCs of heavy and light quarks. 

To summarize, 
in the large-$N$ limit for the projected ENC, the largest angle $R$ is reduced to the \emph{per-jet scalar} observable, and it characterizes the collinear core radius, with soft/csoft radiation parametrically suppressed  by $O(1/N)$. Hence the shape of $d\bar\Sigma/dR$ directly encodes collinear core dynamics of the jet. For heavy flavors, our factorization predicts a robust depletion at small $R$—a one-dimensional manifestation of the dead-cone effect—visible in Fig.~\ref{fig:JENC} and reflected in the $e^+e^-$ study of Fig.~\ref{fig:eplus-eminus}.

\section{Conclusion and Outlook}\label{conc}

In conclusion, we have formulated the factorization theorem for the  ENC in terms of the FFJs, and using the established results of the FFJs, simultaneously resummed the large logarithms of both small $R$ and large $N$. We find that the radius for hard collinear core particles in a jet can be understood as the projected ENC angle $R$ in the large $N$ limit, and its distribution can serve as a key characteristic distinguishing jets of different origins, as shown in the comparison of heavy/light quark ENC.

We expect the framework developed in this work may offer promising applications in collider phenomenology.
In particular, the large-$N$ ENC could serve as a theoretically clean probe of the collinear core structure of jets, with reduced sensitivity to soft contamination, potentially assisting in quark/gluon jet discrimination.
It may also contribute to improved heavy-flavor jet study in highly boosted kinematic regimes and provide complementary observables for identifying multi-pronged substructures arising from boosted heavy objects, such as Higgs bosons or top quarks.
%A more comprehensive phenomenological exploration of these possibilities is left for future studies.
%Moreover, the ENC with large $N$ is expected to play a crucial role in probing quark/gluon jets, boosted Higgs, and other jet phenomena. 
Further rigorous studies, both analytical and numerical~\cite{Budhraja:2024xiq}, are also urgently needed to fully explore these possibilities.

\acknowledgments

LD is supported by the Guangxi Talent Program (“Highland of Innovation Talents”). CK is supported by Basic Science Research Program through the National Research Foundation of Korea (NRF) funded
by the Ministry of Science and ICT (Grant No. NRF-2021R1A2C1008906). AL is supported in part by the National Science Foundation under Grant No. PHY-2412696.

%%%%%%%%%%%%%%%%%%%%%%%%%%%%%%%%%%%%%%%%%%%%%%%%%%%%%%%%%%%%%%%%%%%%%%
%%%%%%%%%%%%%%%%%%%%%%%%%%%%% Bibliography %%%%%%%%%%%%%%%%%%%%%%%%%%%
%%%%%%%%%%%%%%%%%%%%%%%%%%%%%%%%%%%%%%%%%%%%%%%%%%%%%%%%%%%%%%%%%%%%%%

\end{document}